\documentclass[a4paper,11pt]{article}
 \usepackage{amsthm}
 \usepackage{amstext}
 \usepackage{amsmath}
 \usepackage{amsbsy}
 \usepackage{amssymb}
 \usepackage{amsfonts}
 \usepackage{graphicx}
 \usepackage{epsfig}
 \usepackage{authblk}
 \usepackage{subfigure}
 \usepackage{textcomp}
\begin{document}
\title{Steps towards the Hyperfine Splitting measurement of the muonic
 hydrogen ground state: pulsed muon beam and detection system characterization}

\author{
A.~Adamczak$^{a}$,
G.~Baccolo$^{b}$,
D.~Bakalov$^{c}$,
G.~Baldazzi$^{d}$,
R.~Bertoni$^{b}$,
M.~Bonesini$^{b}$,
V.~Bonvicini$^{e}$,
R.~Campana$^{d}$,
R.~Carbone$^{e}$,
T.~Cervi$^{f,g}$,
F.~Chignoli$^{b}$,
M.~Clemenza$^{b}$,
L.~Colace$^{h,i}$,
A.~Curioni$^{b}$,
M.~Danailov$^{e}$,
P.~Danev$^{c}$,
I.~D'Antone$^{d}$,
A.~De~Bari$^{f,g}$,
C.~De~Vecchi$^{g}$,
M.~De~Vincenzi$^{h,j}$,
M.~Furini$^{d}$,
F.~Fuschino$^{d}$,
K.~Gadejisso-Tossou$^{e,k}$,
D.~Guffanti$^{e}$,
A.~Iaciofano$^{h}$,
K.~Ishida$^{l}$,
D.~Iugovaz$^{e}$,
C.~Labanti$^{d}$,
V.~Maggi$^{b}$,
A.~Margotti$^{d}$,
M.~Marisaldi$^{d}$,
R.~Mazza$^{b}$,
S.~Meneghini$^{d}$,
A.~Menegolli$^{f,g}$,
E.~Mocchiutti$^{e}$,
M.~Moretti$^{b}$,
G.~Morgante$^{d}$,
R.~Nard\`o$^{g}$,
M.~Nastasi$^{b}$,
J.~Niemela$^{l}$,
E.~Previtali$^{b}$,
R.~Ramponi$^{m}$,
A.~Rachevski$^{e}$,
L.~P.~Rignanese$^{d}$,
M.~Rossella$^{g}$,
P.L.~Rossi$^{d}$,
F.~Somma$^{h,n}$,
M.~Stoilov$^{c}$,
L.~Stoychev$^{e,k}$,
A.~Tomaselli$^{g,o}$,
L.~Tortora$^{h}$,
A.~Vacchi$^{e,p}$\footnote{Andrea.Vacchi@ts.infn.it},
E.~Vallazza$^{e}$,
G.~Zampa$^{e}$,
and~ M.~Zuffa$^{d}$ \\(FAMU collaboration)}
%\email{Andrea.Vacchi@ts.infn.it}
\affil{$^a$Institute of Nuclear Physics, Polish Academy of Sciences, Radzikowskiego 152, PL31342 Krak\'{o}w, Poland}
\affil{$^b$National Institute for Nuclear Physics (INFN), Sezione di Milano Bicocca, Piazza della Scienza 3, Milano, Italy}
\affil{$^c$Institute for Nuclear Research and Nuclear Energy, Bulgarian Academy of Sciences, blvd. Tsarigradsko ch. 72, Sofia 1142, Bulgaria}
\affil{$^d$National Institute for Nuclear Physics (INFN), Sezione di Bologna, Italy}
\affil{$^e$National Institute for Nuclear Physics (INFN), Sezione di Trieste, via A. Valerio 2, 34127 Trieste, Italy}
\affil{$^f$Department of Physics, University of Pavia, via Bassi 6, Pavia, Italy}
\affil{$^g$National Institute for Nuclear Physics (INFN), Sezione di Pavia, Via Bassi 6, Pavia, Italy}
\affil{$^h$National Institute for Nuclear Physics (INFN), Sezione di Roma Tre, Via della Vasca Navale 84, Roma, Italy}
\affil{$^i$Dipartimento di Ingegneria Universit\`a degli Studi Roma Tre Via V. Volterra, 62}
\affil{$^j$Dipartimento di Matematica e Fisica, Universit\`a di Roma Tre, Via della Vasca Navale 84, Roma, Italy}
\affil{$^k$The Abdus Salam International Centre for Theoretical
 Physics, Strada Costiera 11, Trieste, Italy}
\affil{$^l$RIKEN-RAL}
\affil{$^m$IFN-CNR, Department of Physics - Politecnico di Milano and National Institute for Nuclear Physics (INFN), Sezione di Milano Politecnico, piazza Leonardo da Vinci 32, 20133 Milano, Italy}
\affil{$^n$Dipartimento di Scienze, Università di Roma Tre, Viale G. Marconi 446, Roma, Italy}
\affil{$^o$Department of Electrical, Computer, and Biomedical Engineering, University of Pavia, Via Ferrata 5, Pavia, Italy}
\affil{$^p$Mathematics and Informatics Department, Udine University,
  via delle Scienze 206, Udine}

\maketitle
\begin{abstract}
The high precision measurement of the hyperfine splitting of the muonic-hydrogen atom
 ground state with pulsed and intense muon beam requires careful
 technological choices both in the construction of a gas target and of the
 detectors. In June 2014, the pressurized gas target of the FAMU
 experiment was exposed to the low energy pulsed muon beam at the
 RIKEN RAL muon facility. The objectives of the test were the characterization
 of the target, the hodoscope and the X-ray detectors.

The apparatus consisted of a beam hodoscope and X-rays detectors made with high purity Germanium and Lanthanum Bromide crystals. In this paper the experimental setup is described and the results of the detector characterization are presented.
\end{abstract}

%\keywords{Hyperfine splitting; Zemach radius; Muonic hydrogen}

%\collaboration[c]{(FAMU collaboration)}

\newpage

\section{The FAMU project}

The FAMU (Fisica degli Atomi MUonici) project is an experimental
 program for the determination of the Zemach radius r$_Z$ of the proton, the only observable parameter that characterizes both its charge and magnetic distributions. To this end, the hyperfine splitting (HFS) in the muonic-hydrogen atom ground state will be measured~\cite{bakalov92,adamczak12,bakalov15}. It is a science-driven, technologically ambitious project motivated by the controversy about the proton size, that calls for new approach to proton's structure studies. The published measurement of the proton rms charge radius, extracted from the experimental value of the Lamb shift in muonic hydrogen, differs by $9\sigma$ from previous data derived from hydrogen spectroscopy and electron scattering experiments \cite{bib1,bib2}. Discussions on the correctness of the experiment or models' adequacy have not lead to an explanation of this discrepancy and do not exclude anomalies within the electron-muon universality.

In this experiment a pulsed high-intensity muon beam entering a hydrogen gas target will form muonic hydrogen atoms. The hyperfine splitting in the ground state of the muonic hydrogen atom will be measured by stimulating spin-flip transitions with a pulsed mid-infrared (MIR) tunable laser and counting the number of muon transfer events N($\lambda$) to a higher-Z gas admixture in the hydrogen target as function of the laser wavelength $\lambda$; the resonance laser wavelength $\lambda_0$ is recognized by the maximal response of N($\lambda$) to variations of $\lambda$. The method is based on the following chain of processes. A muonic hydrogen atom in the ground para (F=0) state, after absorbing a photon of the hyperfine splitting resonance energy $\Delta E_{HFS} \sim$ 0.182 eV, and being excited to the ortho (F=1) spin state, is very quickly de-excited in subsequent collisions with the surrounding H$_2$ molecules. At the exit of the collision the muonic atom is accelerated by $\sim$2/3 of the excitation energy $\Delta E_{HFS}$, which it takes away as kinetic energy. In subsequent collisions the muon is transferred to a higher-Z nucleus. The number of muon transfer events depends on the number of spin-flip events, provided that the rate of muon transfer depends on the collision energy. Monte Carlo simulations have demonstrated the efficiency of this method assuming the availability of MIR tunable lasers and pulsed muon sources with sufficient intensity, and the existence of gases for which the muon transfer rate displays significant energy dependence in the epithermic range. The theoretical and experimental effort will face the proton charge radius puzzle by strengthening the experimental limits on the proton structure parameters and measuring the muonic atom HFS with unprecedented precision - $\frac{\delta\lambda}{\lambda}< 10^{-5}$  - and will shed light on the low momentum limit of the magnetic-to-charge form factor ratio. 

The FAMU project foresees a progressive approach to the final measurement of the 1S state hyperfine transition on the muonic hydrogen atom. In the first phase a preliminary experimental layout is set up to perform measurements of the collision energy dependence of the $\mu ^-$  transfer rate in various gases. In particular, the objective of the first phase of the experiment is a complete study of the muon transfer to gases for which there are experimental evidences or theoretical hints for a pronounced energy dependence of the transfer rate in the epithermal range. The results will set the required firm ground for all following activities and will be used to determine the optimal temperature, pressure, laser shot timing and the chemical composition (admixture gas and its concentration) of the gas target for the measurement of the HFS in muonic hydrogen.

\section{The experimental setup}
This paper refers to the preparatory test experiment \cite{R512} which was realized on the pulsed muon beam of the RIKEN-RAL muon facility \cite{RAL} at Rutherford Appleton Laboratories. The main goals are to test and characterize the detection system on a muon beam line and to study the properties of the gas target pressurized vessel.

At the RIKEN-RAL facility, the negative muon beam has a double pulse structure with 70 ns pulse width (FWHM) and 320 ns peak to peak time distance with a 50 Hz repetition rate. The intensity is about 8$\cdot 10^4$ $\mu^-$/s at 60 MeV/c, the energy spread is $\sim$ 10\% and the angular divergency is 60 mrad. The beam spatial profile, without collimator, is circular with a size of $\sigma_r \sim$ 4 cm. The trigger signal for the experiment was given by the beam line itself.

The experimental layout consisted of:
\begin{itemize}
 \item a fiber-SiPM hodoscope;
 \item a high pressure gas target;
 \item two high purity Germanium detectors (HPGe);
 \item a 1$\times$1 inches lanthanum bromide (LaBr$_3$) detector;
 \item a LaBr$_3$(Ce) mosaic, composed by four 0.5$\times$0.5 inches crystals.
\end{itemize}

\begin{figure}[!tbh]
\centering
\includegraphics[width=0.95\textwidth]{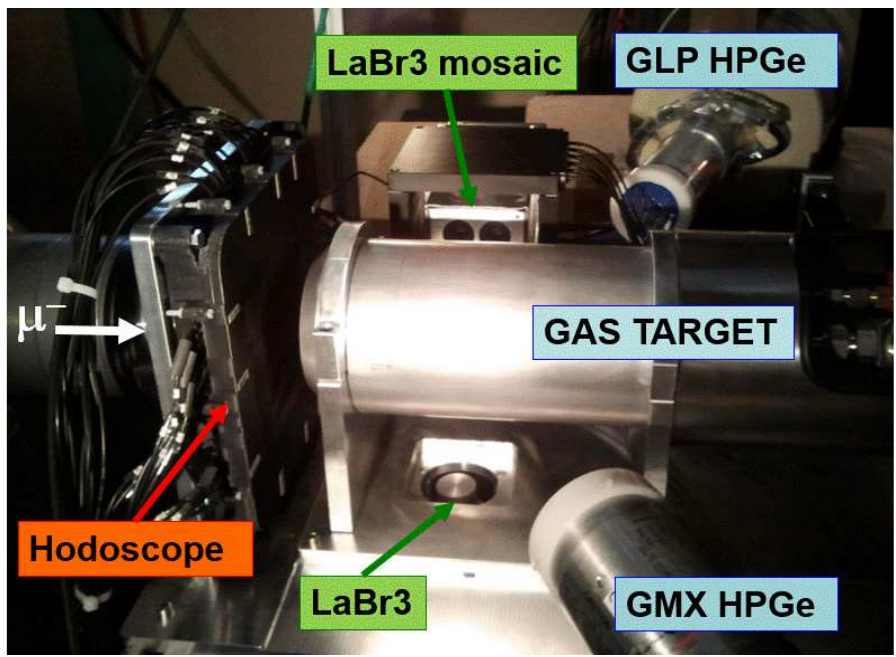}
\caption{Side view of the experimental set-up in operations at RIKEN-RAL.}
\label{fig:1}       
\end{figure}

Detectors were arranged as shown in Fig. \ref{fig:1}.
The aluminium vessel of the gas target can be seen in the middle of the picture, placed on the beam line. On the left of the picture the beam collimator is visible; moving clockwise it is possible to notice the box containing the LaBr$_3$ mosaic and the two HPGe detectors. The 1 inch LaBr$_3$ detector is also visible in the hole next-to the target, positioned under the experimental table and facing the target itself from below at a distance of about two cm. The beam hodoscope is placed between the collimator and the gas target. The distance between the beam axis and X-ray detectors is about 15 cm.

\subsection{The fiber-SiPM hodoscope}
\begin{figure}[!tb]
\centering
\includegraphics[width=0.95\textwidth]{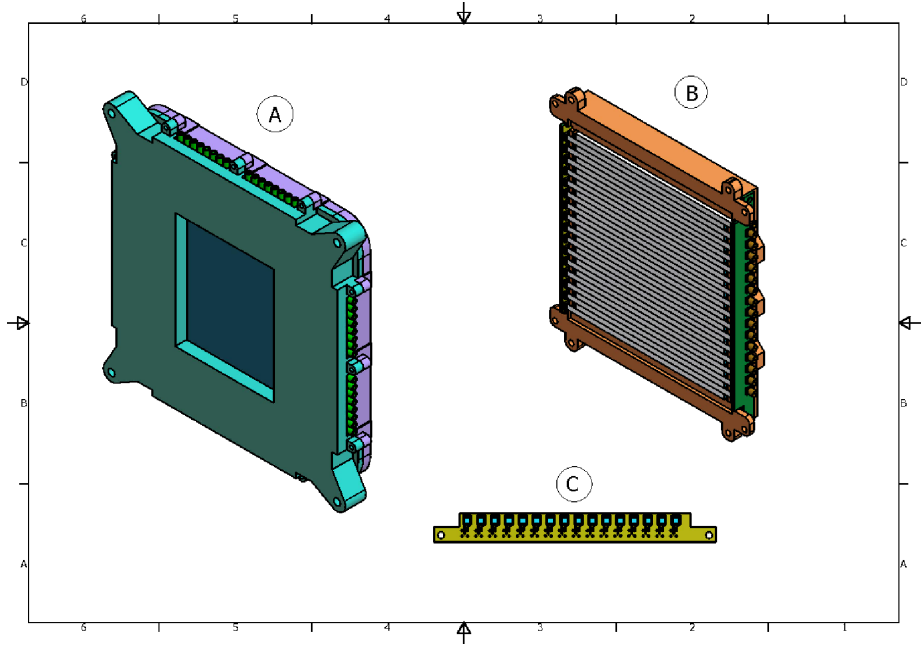}
\caption{Mechanics layout of the beam hodoscope. A) is the general structure of the light-tight assembly. Signal/power cable are fanned out through the feed trough along the lateral side of the structure. B) shows one of the two holders for the scintillating fibers (in this case  along x) and read alternatively at one edge by SiPM mounted on custom PCBs. MCX connectors are shown. C) layout of one PCB with the mounted SiPMs, on the other side MCX connectors convey signal and power for each individual channel.}
\label{hodo}       
\end{figure}

The hodoscope is made by scintillating fibers read with SiPMs \cite{carbone2015}, see Fig. \ref{hodo} for details.
 It is designed to measure the position, shape and timing of the low energy muon beam entering the gas target. Its very compact design is made of two arrays of 3~mm square shaped Bicron BCF12 scintillating fibers, along x and y coordinates, arranged in planes with active area around 10 x 10 cm$^2$. Each fiber is read at one end by a 3$\times$3 mm$^2$ RGB Advansid SiPM\footnote{breakdown voltage $\sim$29~V, 50 micron cell size}, for a total of 64 read-out channels. 
 The SiPM's are mounted in groups of 16 on 4 custom made PCB's (see figure 2 for details) facing alternatively on the two sides the 32 fibers of a x(y) detector plane. For space problems (i.e. the SiPM' footprint as compared to the fiber size) fibers are read from one edge only, alternating up/down or left/right sides. The detector is housed in a 3D printed ABS case. All the system  is  powered and controlled by means of custom made electronics, developed as part of the TPS project~\cite{branchini2011}. Each SiPM is connected via a single RG174 cable to the TPS board, conveying both SiPM's powering (on the external braid shield) and signal output (on the cable inner conductor. Each RG174 cable has MMCX connectors on the PCB's side and Lemo 00 connectors on the TPS side. Each module (8 channels) provides the fine regulation of the individual SiPM voltages, signal amplification and shaping, signal discrimination and trigger capability. The individual output signal have then been fed into a CAEN V792 QADC for measurement of the charge integrated signal.

\begin{figure}[!tb]
\centering
\includegraphics[width=0.95\textwidth]{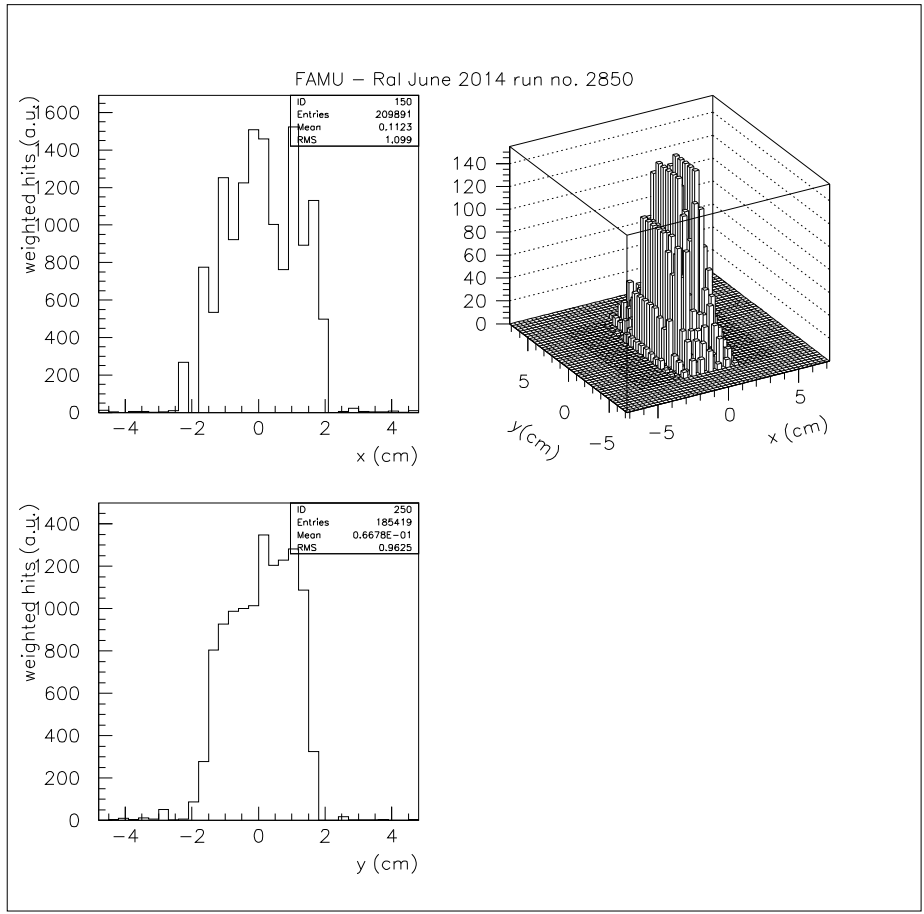}
\caption{Weighted beam profile for a typical run at 61MeV/c. Left panels (top/bottom): x/y projections of beam profile. Right panel: beam spot as a x-y lego plot.}
\label{hodo2}           
\end{figure}
Fig. \ref{hodo2} shows the beam profile obtained from the fiber-SiPMT
 hodoscope for a run at 61 MeV/c. Entries are weighted with QADC
 response, thus giving a beam profile in aribrary units. From the
 figure it can be noticed how the beam has a 1.10 (0.97) cm r.m.s in x(y) coordinates.

\subsection{The gas target}
The core of the apparatus is the gas target. Muons from the beam line
 enter and stop in the gas creating excited muonic atoms. The
 de-excitation generates X-rays that can be detected once they exit the
 target. Muons eventually are transferred to high Z atoms, decay or undergo the nuclear capture, producing other secondary particles (electrons, photons, neutrons and neutrinos) which can contribute to the background.

The target is made of an aluminium cylindrical vessel filled with high pressure gas. The vessel, constructed by Criotec Impianti S.r.l. \cite{WEBCrio}, is built using the Aluminum alloy Al6061. It is a cylinder of 125 mm of base diameter and 260 mm length, with an inner volume of 2.8~l. The thickness of the walls is 7 mm, except for a circular entrance window of 44 mm diameter thinned to 4 mm. This shape and dimensions ensure a perfect and certified resistance to gas pressure of several tenth of bar and minimize the divergence of the muon beam due to multiple scattering.

Three gas mixtures were used during the test beam: a pure (99.999$\%$) hydrogen gas, a gas mixture of 4$\%$($\pm 0.08\%$) carbon dioxide (CO$_2$) in hydrogen and a gas mixture of 2$\%$($\pm 0.04\%$) argon in hydrogen. All used mixtures were made by weight. The filling procedure was performed by first evacuating the gas target to a partial pressure of 10$^{-6}$ bar and then filling it with the chosen high purity gas.

\subsection{The HPGe X-ray detectors}
The ORTEC GLP Series Planar Low-Energy Photon Spectrometer (LEPS) is a small-area, high-purity germanium photon spectrometer for use in applications over a typical energy range from 3 to 300 keV. The LEPS offers exceptional energy resolution for low and intermediate energies. This detector is located on the left side of the target, see Fig. \ref{fig:1}. It is a cylindrical crystal with diameter of 11 mm and 7 mm height. The entrance window is made by a beryllium layer of 127 $\mu$m and the inactive germanium depth is 0.3 $\mu$m.
 The FWHM energy resolution at 5.9 keV is 0.2 keV and at 122 keV is 0.5 keV using a 6~$\mu$s gaussian pulse shape with a ORTEC 672 amplifier.

The ORTEC GMX Series HPGe Detector is a N-type Coaxial HPGe Detector with a thin entrance window made by a beryllium layer of 0.5 mm and a outside contact layer of 0.5 $\mu$m of inactive Ge with implanted Boron Ions. The type used in this work is a 20$\%$ relative efficiency detector in pop-top configuration with a crystal diameter of 54.8 mm and 49.8 mm length. The FWHM energy resolution at 5.9 keV is 0.6 keV and at 122 keV is 0.8 keV using a 6~$\mu$s gaussian pulse shape with a ORTEC 672 amplifier.

\subsection{The LaBr$_3$(Ce) X-ray detectors}
The LaBr$_3$(Ce) is a crystal that offers the best energy resolution among scintillators, fast emission and excellent linearity. 
 The primary decay time is 16 ns and the coincidence resolution time is better than 1 ns. The energy resolution is located halfway between scintillators and solid state detectors (3\% at 662 keV, 6.5\% at 122 keV). The light output is 160\% higher respect to NaI(Tl) with maximum emission at 380 nm. A significant deviation from linearity in light emission is seen only at a rate of over a million events per second \cite{LABR0, LABR1,LABR2}. A radioactive background is present in LaBr$_3$(Ce), mostly due to $^{138}$La isotope. This isotope decays into Barium and Cesium emitting X-rays of 1436 and 789 keV \cite{LABR3}. An accurate determination of the background (by self-counting the spectrum) gives (0.077 $\pm$ 0.004) cps/g \cite{LABR2,LABR4}. The number of spurious events due to the background generated by the crystal itself is therefore negligible, considering the crystal dimensions and the acquiring time window of 5000 ns. Two set of lanthanum bromide detectors were used.

The first one is a commercially available BrilLanCe$^{TM}$ 380 detector by Saint-Gobain Crystals \cite{LABR2}. It is provided in a mounted assembly based on a cylindrical 1 inch diameter 1 inch long LaBr$_3$(5\% Ce) crystal coupled to a XP2060 photomultiplier.

\begin{figure}[!tb]
\centering
\includegraphics[width=0.59\textwidth]{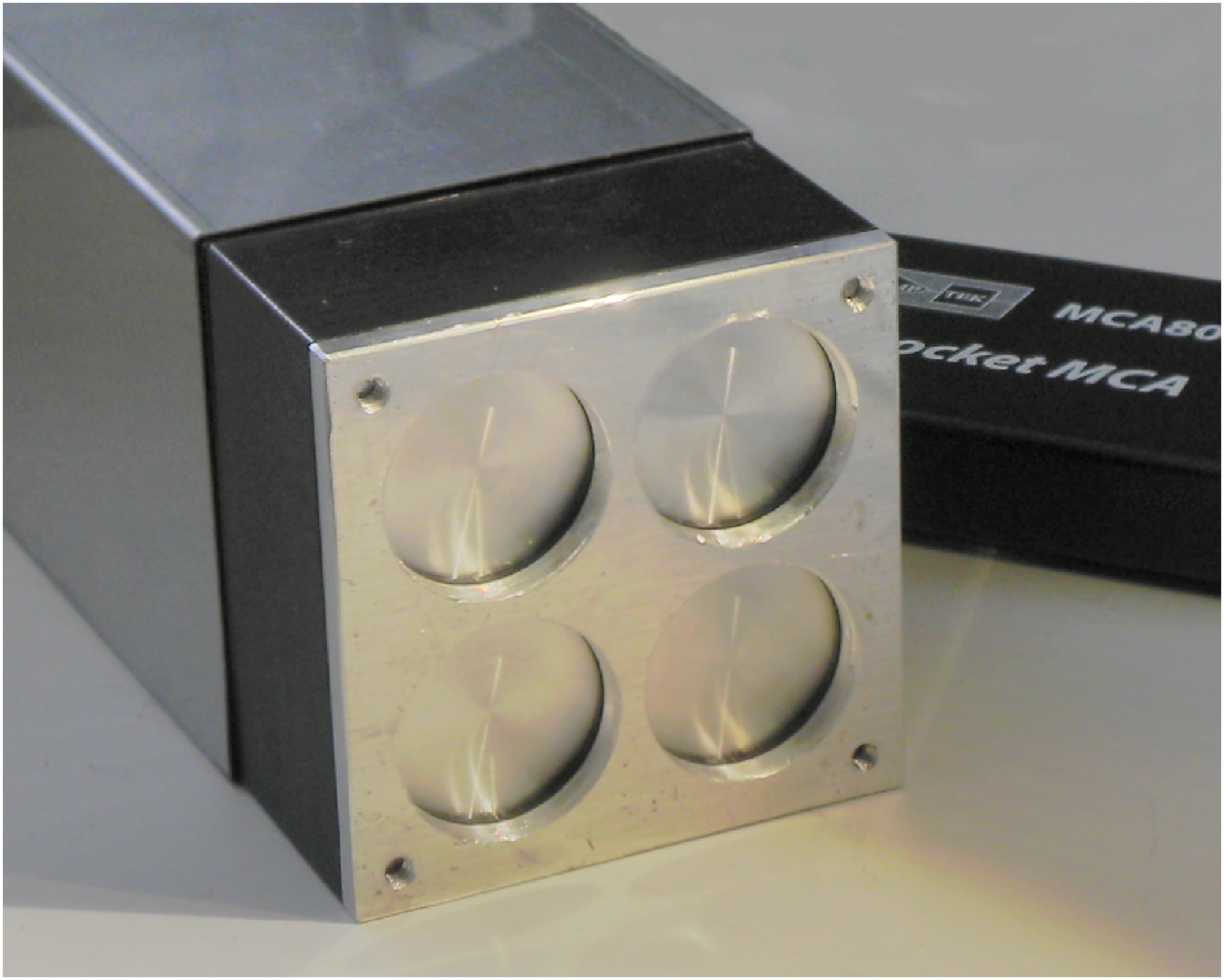}
\includegraphics[width=0.39\textwidth]{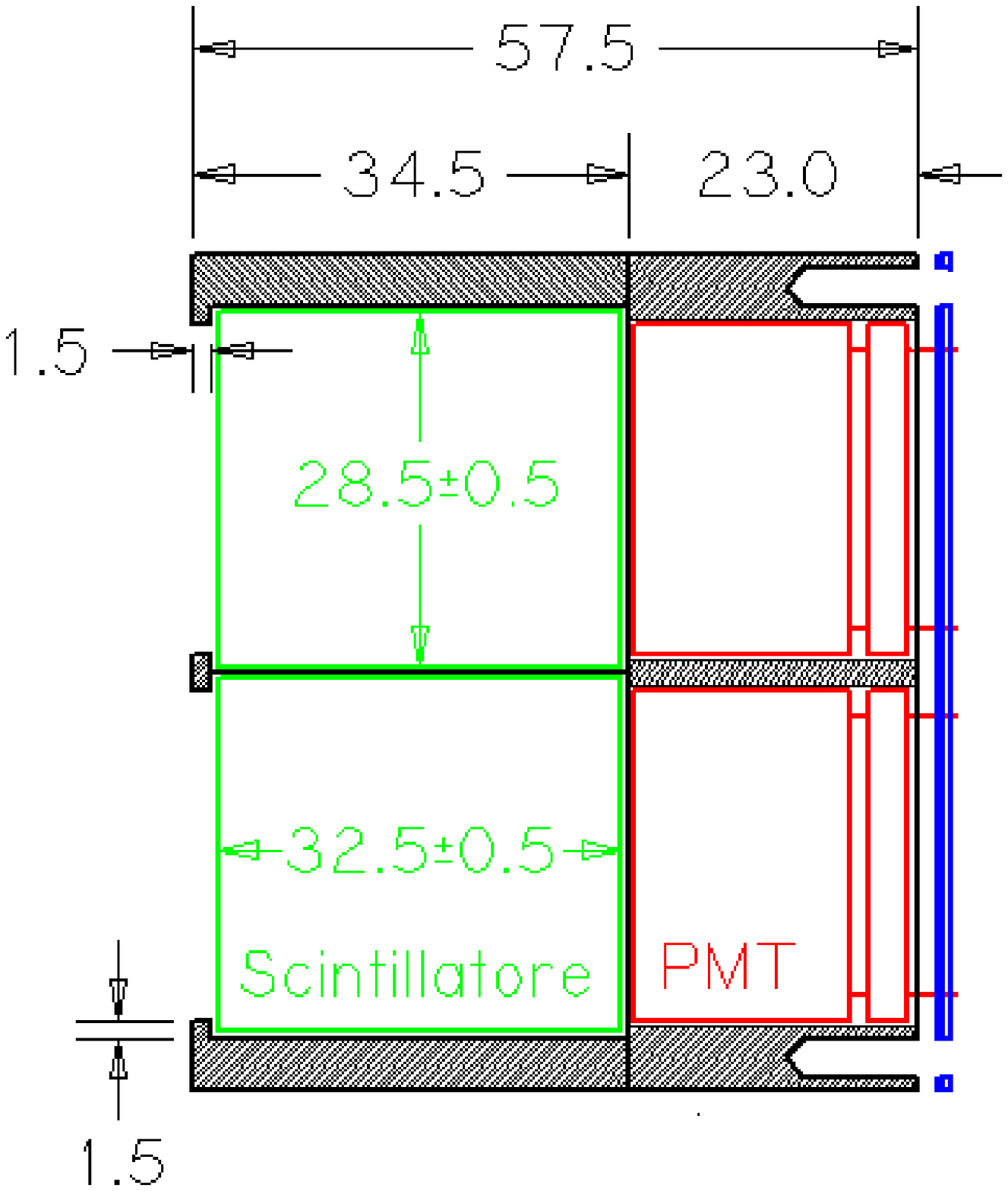}
\caption{Left: picture of the detection module based on a mosaic of LaBr$_3$(Ce) crystals. Right: drawing of the LaBr$_{3}$+PMT systems composing the mosaic detector (side view). The dark grey block, light-tight, was made with a 3D printer. All dimensions are expressed in mm. }
\label{fig:labr_pic}       
\end{figure}
The second detector is a mosaic of four smaller LaBr$_3$(Ce) crystals, 0.5 inches diameter 0.5 inches long, coupled to Hamamatsu R11265-200 high speed, ultra bialkali photomultipliers (see picture in Fig. \ref{fig:labr_pic}). The four detectors have been placed in a 2x2 matrix, as schematically reported in figure \ref{fig:labr_pic} on the right, to form the mosaic, with crystals facing the target, and housed in a 80x80x200 mm iron box coated, on side, by a 2 mm deep lead sheet. PMTs were equipped with a modified HV divider to read high charge pulse produced by LaBr$_3$(Ce).

\section{Data acquisition and processing}
Signals from all detectors (except the hodoscope) were digitized at 500 MHz, using a CAEN D5730 digitizer, and recorded; the 64 hodoscope output channels were read by two CAEN V792 QADC and recorded in time.

The trigger signal was given by the beam line and the acquisition started about 250 ns before the time at which the center of the first muon pulse reached the target. Then the digitizer samples, for each detector, the output signal every 2 ns in a chosen time window of about 5~$\mu$s. The advantage of using the digitizer lies in such data sampling: by recording an \textquotedblleft oscilloscope-like\textquotedblright output for each detector and each trigger event, and for a time much longer than the typical muon transfer, a complete view of all features occurring during the measurement is obtained. Hence there is a powerful chance to identify and study peculiarities of photon and other particles signals, noise structures and time decay. Moreover, avoiding the processing of the signal by amplifier and shaper stages, a cleaner and faster signal timing can be recorded. 

For each trigger the syncronized output of all channels from digitizer and QDC was recorded on disk. Sets of data from 15000 trigger events were acquired as a single run and stored in n-tuples. All measurements contained in a single run were performed in the same experimental conditions.

During 5 days of experimental operations about 800 runs have been recorded, for a total of more than 10$^7$ triggered events:
\begin{itemize}
 \item 150 runs were taken to test and calibrate detectors and electronics;
 \item 90 runs were taken using a graphite target;
 \item 50 runs were taken with pure gaseous hydrogen (H$_2$) target at 35 bar pressure;
 \item 350 runs were taken with H$_2$+CO$_2$ (4\%) 38 bar target (50 of which by varying the beam kinetic energy in the interval 59-63 MeV);
 \item 120 runs were taken with H$_2$+Ar(2\%) 40 bar target.
\end{itemize}
 The graphite target represents the simplest case of muonic lines production. This measurements were used to calibrate the detector system.

The pure hydrogen target measurements were needed to set the background generated by the aluminium container both in the energy and time spectra. In fact, the results of measurements with contaminant gas were compared to the pure hydrogen ones in order to study the performances and capabilities of the whole setup in studying the muon capture and transfer mechanisms induced by oxygen and argon.

\section{Optimization of the kinetic energy of the muon beam}
The kinetic energy of the muon beam was tuned in order to maximize the stop of muons in the gas target. If the kinetic energy is too low muons will stop in the entrance Aluminum window while if it is too high muons will exit the target without stopping. The detectors response was monitored online to determine the height of the K$\alpha$ spectral line for the muonic oxygen in the H$_2$+CO$_2$ gas mixture. The HPGe detector was used for this purpose while the muon beam momentum was increased from 59 to 63 MeV/c by 1 MeV/c step (corresponding to the finest possible tuning). At each energy step the same number of trigger events was recorded. The number of events for the muonic oxygen K$\alpha$ spectral line was determined. The maximum of the muon stop in the gas corresponds to the momentum value of 61 MeV/c which was chosen as preferable momentum for all the other gas measurements. 

\section{Data analysis}
The aim of the 2014 beam test was to determine the suitability of the previously described detectors for the proposed measurement. In an environment with a possible large background, the detection of characteristic X-rays was not taken for granted. Hence the search of these X-rays lines was performed initially by using a simple graphite target (pure carbon) and then by using a complex system like an aluminium vessel filled with several mixtures of hydrogen and other gases.

The first task of the analysis was to identify the peaks due to the characteristic X-rays of the muonic atoms in the energy spectrum obtained by the HPGe and the LaBr$_3$ detectors.

Then, the event time distribution was studied to determine the muon-transfer speed from hydrogen to heavier gases. The time spectrum is particularly important for the H$_2$+CO$_2$ mixture, since oxygen is the main candidate for a muon-transfer experiment given the energy dependence of the transfer rate at epithermal energies~\cite{werthmuller96}.

As already mentioned in previous section, for each X-ray detector and for each trigger event a complete 5 $\mu$s dump of the digitized detectors waveforms was stored on disk. The first step of the analysis was to filter each dump to identify photon signals over the background. Time of arrival and energy have been reconstructed for each photon peak by fitting the pulse; in case of pile-up a multiple fitting procedure has been applied to correctly determine the desired parameters.

\subsection{Energy spectra}
The first target to be considered was a small high-purity $\sim$1cm thick graphite block. From a block of such a pure material, only the characteristic X-rays of carbon are expected. The relevant transitions in the muonic carbon atom ($\mu$C) are \cite{daniele72} the K$_\alpha$ (75.2588 keV), K$_\beta$ (89.212 keV) and K$_\gamma$
(94.095 keV) lines.
\begin{figure}[!tb]
\centering
\includegraphics[width=0.95\textwidth]{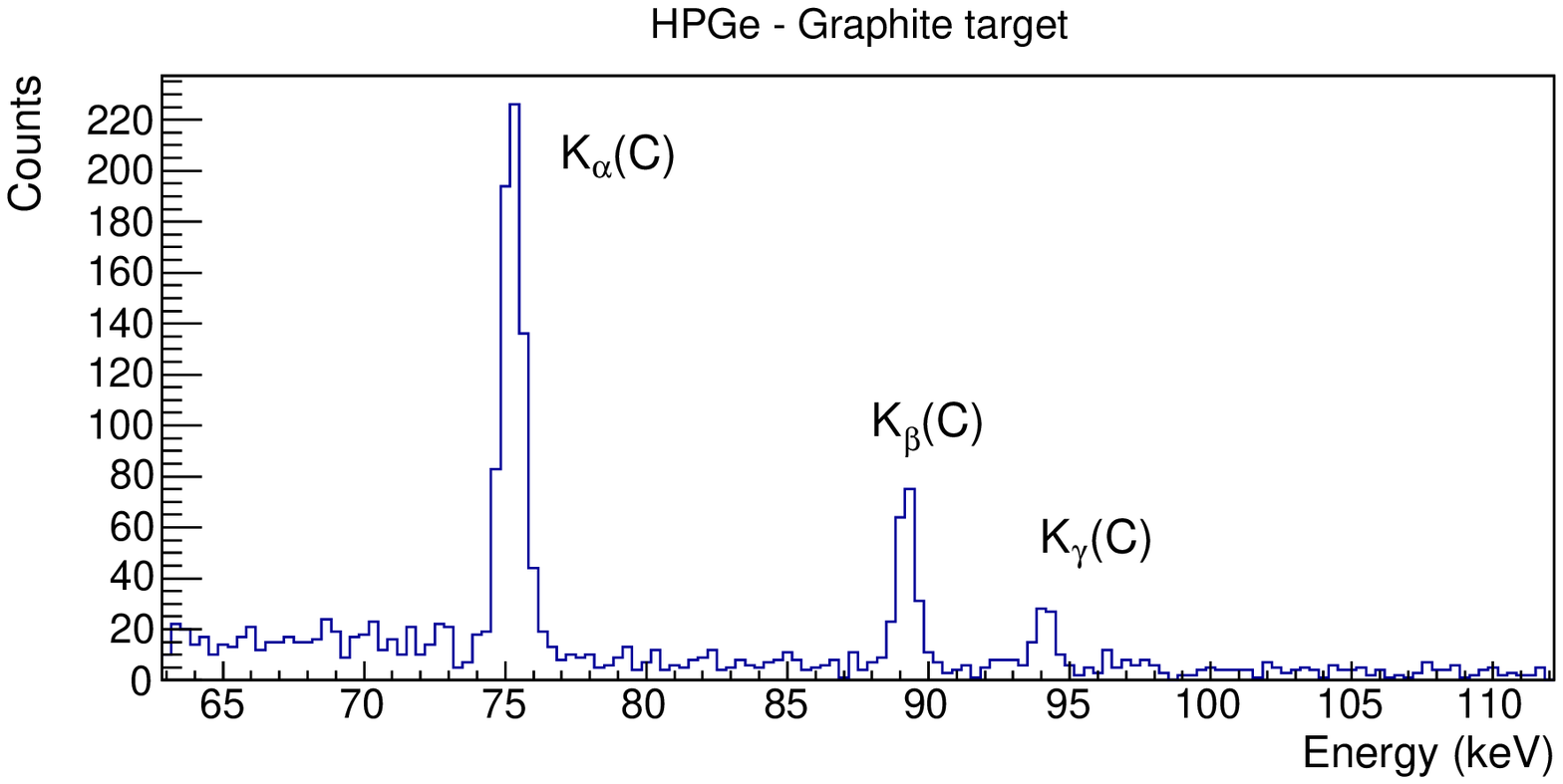}
\includegraphics[width=0.95\textwidth]{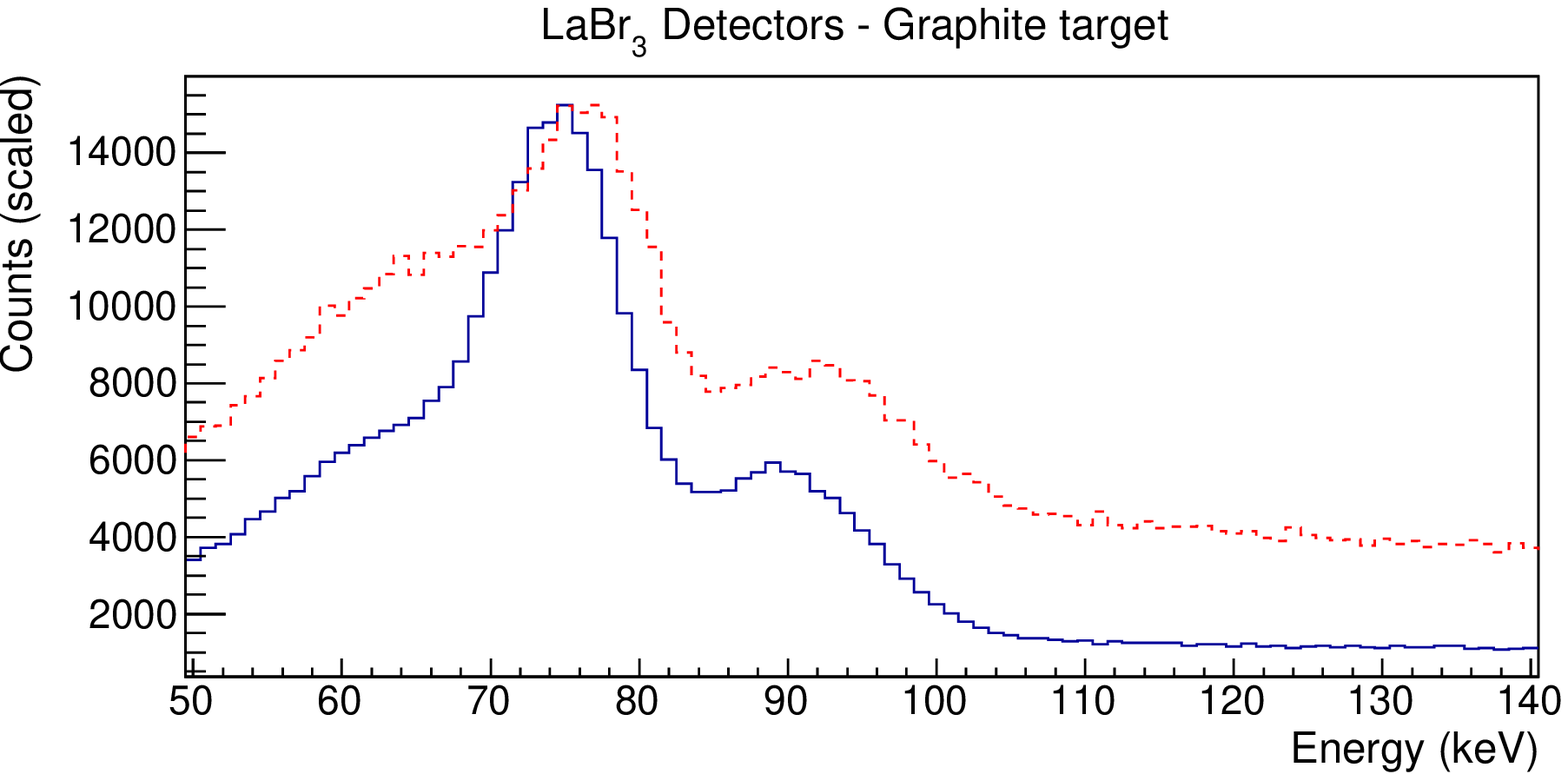}
\caption{Graphite target energy spectrum as measured by: the GLP HPGe
  detector (top panel), the 1$\times$1 inches LaBr$_3$ detector
  (bottom panel red dashed line), the LaBr$_3$ mosaic detector (bottom
  panel blue solid line). K$_\alpha$, K$_\beta$ and K$_\gamma$ lines of $\mu$C are visible.}
\label{fig:graphite_en}       
\end{figure}
 The purity of the target was confirmed by the X-ray spectrum measured by the HPGe detector, where all the lines can be ascribed to muonic carbon, see top panel of Fig. \ref{fig:graphite_en}. The X-ray spectrum recorded by the LaBr$_3$ detectors is shown in Fig. \ref{fig:graphite_en} bottom panel. The number of event acquired with these detectors was much greater than for the HPGe detector. As expected, the energy resolution for LaBr$_3$ detectors can not compete with the germanium detector. However, the resolution is enough to resolve the main peak at 75.26 keV from the ones at 89.21 and 94.1 keV.

Among the different gaseous targets, the simplest one is a pure hydrogen target. While muonic hydrogen transitions are not expected to give any contribution in the X-ray spectrum (in the energy region above tens of keV), the aluminium of the vessel must be considered. In fact, a large number of muons is stopped in the walls of the aluminium vessel where they form $\mu$Al atoms. A study of a pure H$_2$ target is thus needed in order to understand the background due to aluminium characteristic X-rays. In the X-ray spectrum obtained by the HPGe detector 
\begin{figure}[!tb] 
\centering 
\includegraphics[width=0.95\textwidth]{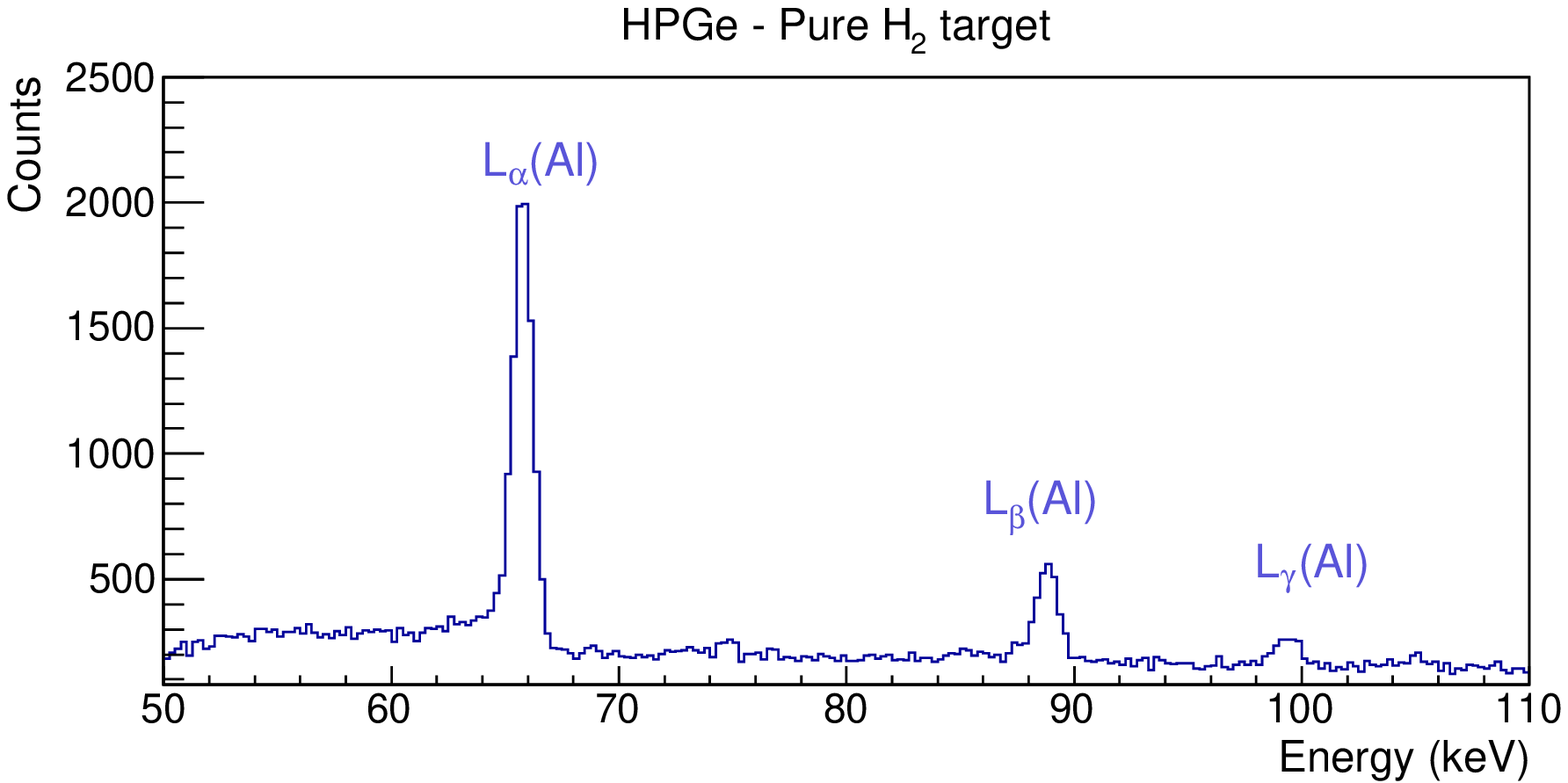} 
\includegraphics[width=0.95\textwidth]{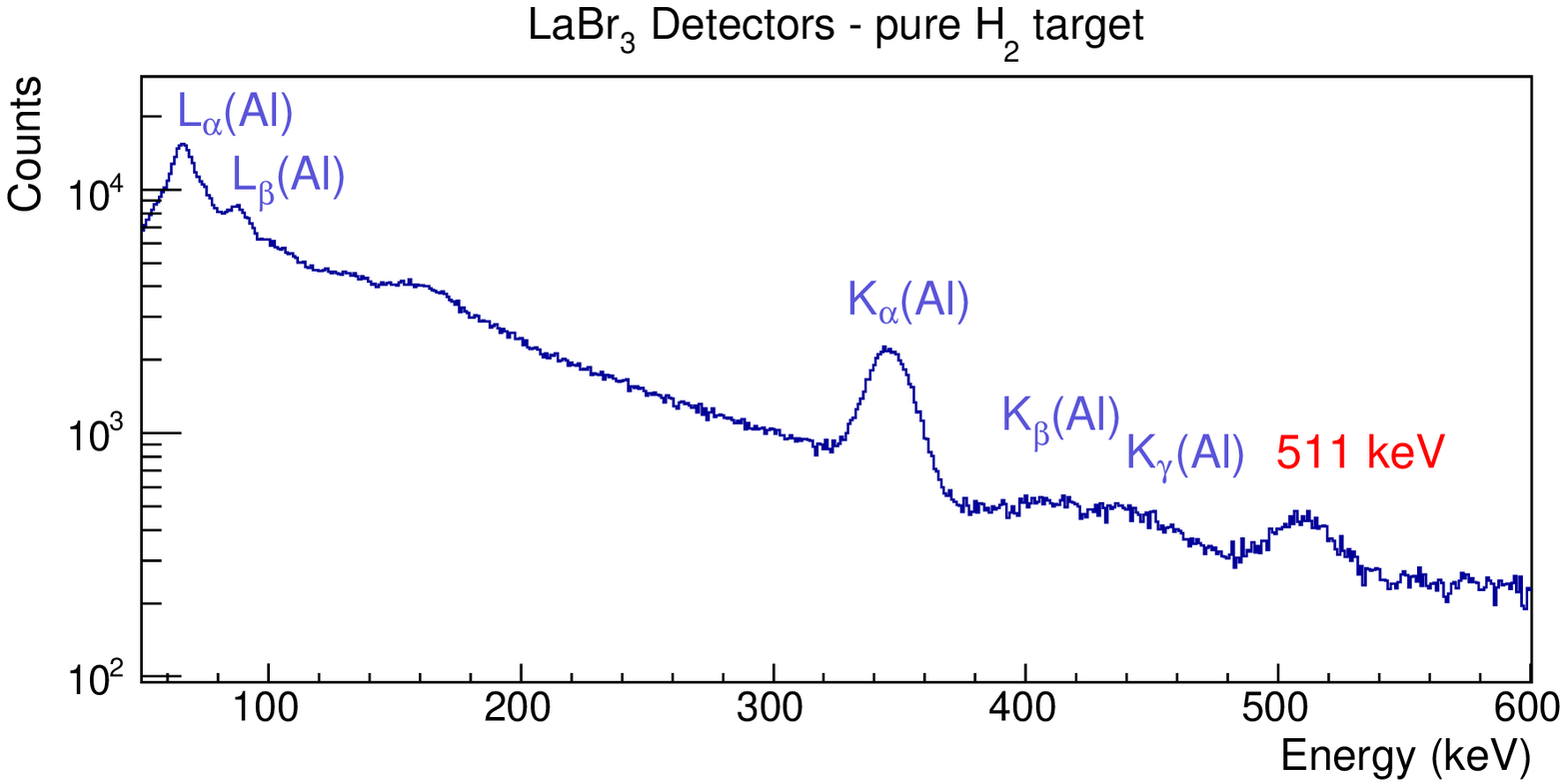} 
\caption{Top panel: energy spectrum as measured by the GLP HPGe detector with the pure H$_2$ gas target. L$_{\alpha}$, L$_{\beta}$ and L$_{\gamma}$ lines of $\mu$Al are visible. Bottom panel: energy spectrum as measured by the LaBr$_3$ mosaic with the pure H$_2$ gas target.} \label{fig:h2_en} 
\end{figure} 
 (Fig. \ref{fig:h2_en} top panel) only the characteristic X-rays of aluminium are present. This is an evidence that the gasoeus target is sufficiently clean. The X-ray spectrum obtained using the LaBr$_3$ scintillating detectors (Fig. \ref{fig:h2_en}) confirms the observation of the characteristic lines of aluminium, including the K-lines (347, 413 and 436 keV).

A mixture of hydrogen and argon (H$_2$+2$\%$Ar) was also tested. The spectrum obtained from the LaBr$_3$ scintillating detector allows to detect also the argon K$_{\alpha}$ transition at 673 keV 
\begin{figure}[!tb]
\centering
\includegraphics[width=0.95\textwidth]{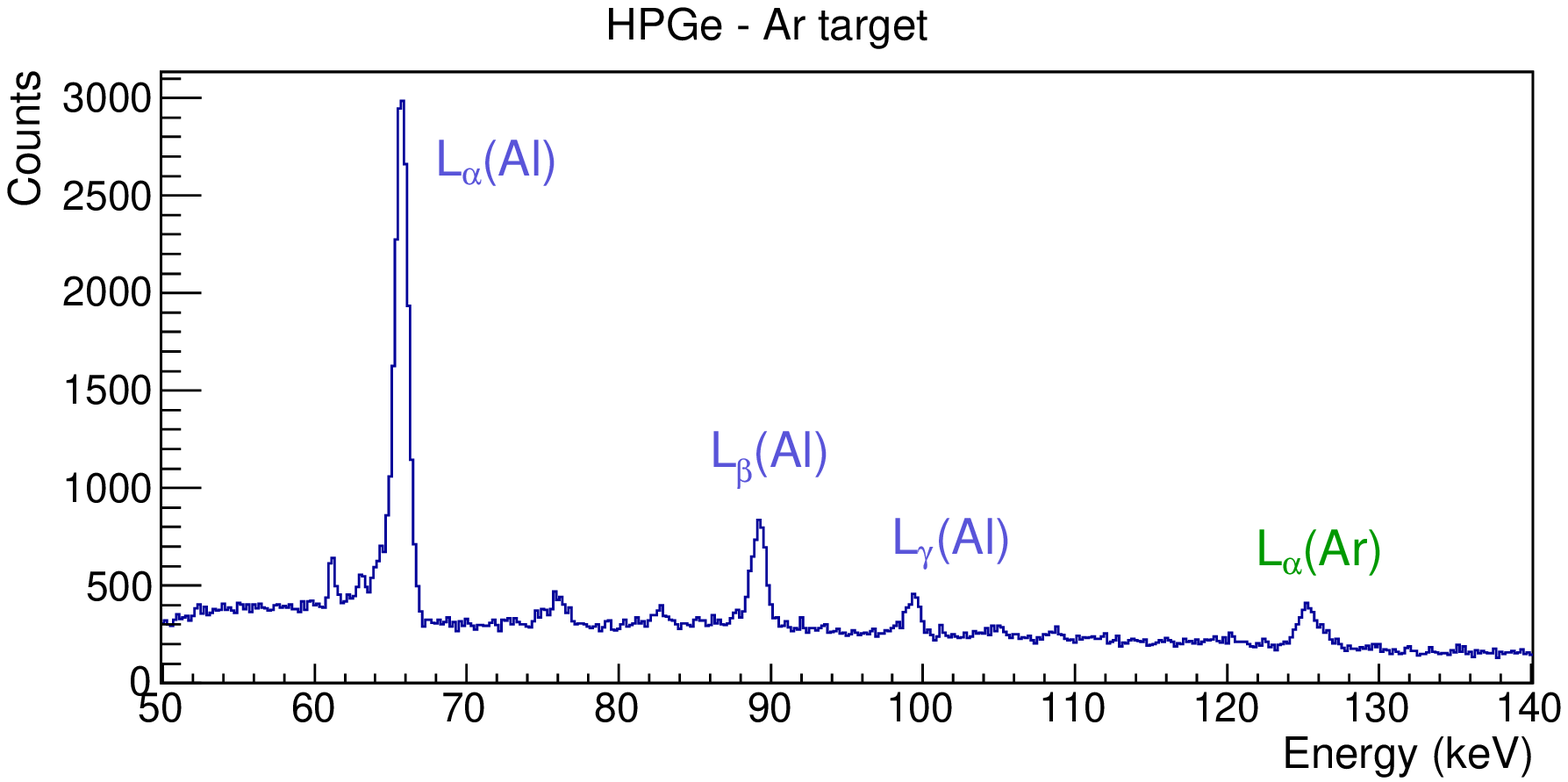}
\includegraphics[width=0.95\textwidth]{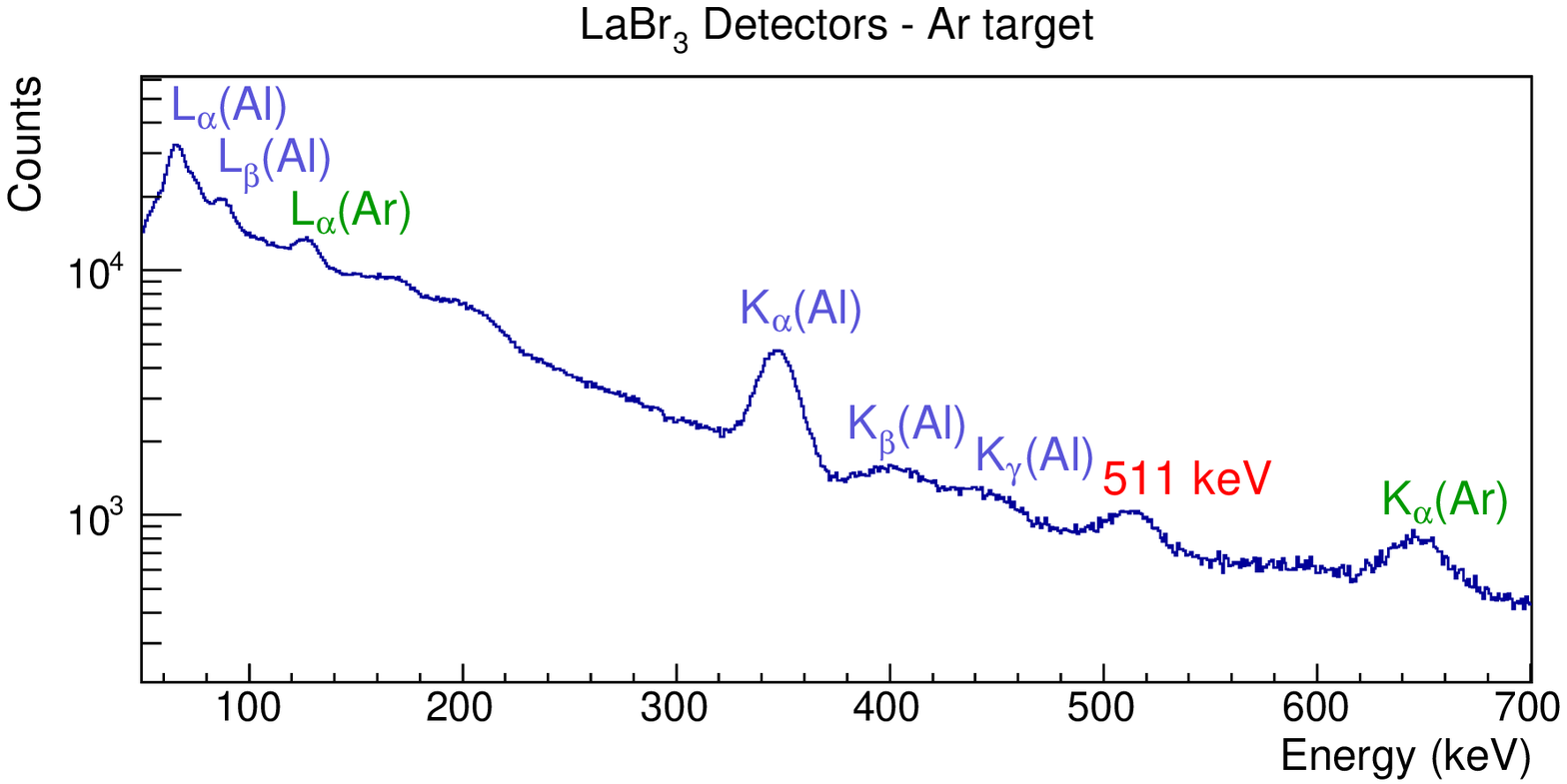}
\caption{Energy spectrum as measured by the GMX HPGe detector with the H$_2$+2$\%$Ar gas target: the K$_\alpha$ line of $\mu$Ar arises from the background. Energy spectrum as measured by the LaBr$_3$ mosaic with the H$_2$+2$\%$Ar gas target.}
\label{fig:ar_en}       
\end{figure}
 (Fig. \ref{fig:ar_en}, bottom panel). In the energy spectrum recorded by the GLP HPGe detector, same figure top panel, only the aluminium lines together with the argon L$_{\alpha}$ peak are present, confirming the purity of the gas mixture.

As mentioned earlier, the most promising candidate for the gas mixture is oxygen. In fact, oxygen is one of the few elements that showed a clear energy dependence of the muon-transfer rate in the epithermal range, and for this reason some experimental studies have already been performed in the nineties~\cite{werthmuller96,schnu92,mulhauser93,werthmuller98}. The natural choice for testing the detectors sensitivity to oxygen and the presence of an anomalous muon-transfer rate at epithermal energies would have been to use a H$_2$+O$_2$ mixture. However, even low concentration of oxygen in a mixture with hydrogen at high pressure is a problem from the safety point of view, given the possibility of explosion. This problem can be solved using carbon dioxide (CO$_2$) instead of molecular oxygen in the gas mixture; thus the target was filled with a mixture of 96\% H$_2$ and 4\% (mass) CO$_2$ at a pressure of 38 bar, which, in terms of atomic concentration of oxygen, results in $c_O \sim 0.19$\%.

This concentration was chosen conservatively in order to observe the
 $\mu$O transition lines. However, it is too high to grant sufficiently delayed muon-transfer events to
 oxygen \cite{bakalov15}. Hence the delay of oxygen X-rays due to the muon-transfer rate from $\mu$p to oxygen at epithermal energies is not expected to be particularly relevant. Once again, the energy spectrum measured by the GLP HPGe detector, see
\begin{figure}[!tb]
\centering
\includegraphics[width=0.975\textwidth]{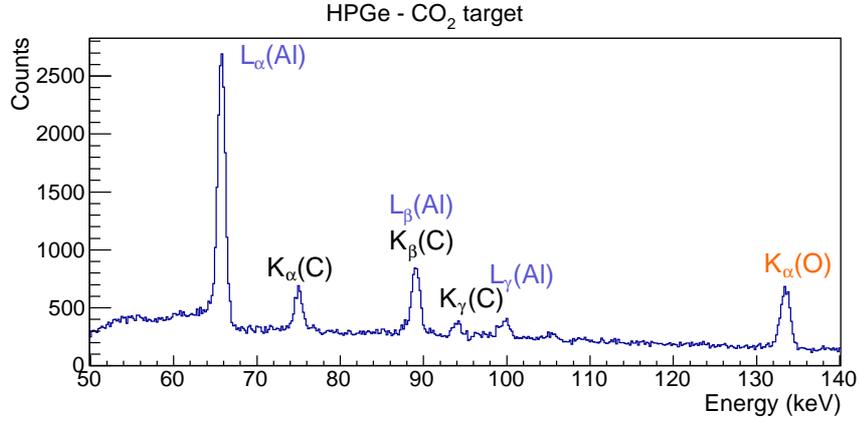}
\caption{Energy spectrum as measured by the GLP HPGe detector with the H$_2$+4$\%$CO$_2$ gas target in the region of K spectral lines from $\mu$Al to $\mu$O. Detected spectral lines are labeled. From this comparison it can be noticed that there was not gas contamination and all measured events in $\mu$O spectral lines originated in the gas target.}
\label{fig:co2_hpge}       
\end{figure}
 Fig. \ref{fig:co2_hpge}, was used to guarantee that there were no contamination in the gas target and the observation of the lines corresponding to the $\mu$Al, $\mu$C and $\mu$O transitions confirms the composition of the gas mixture. The energy spectrum as recorded by the LaBr$_3$ detectors is shown, instead, in Fig. \ref{fig:co2_labrmi}.
\begin{figure}[!tb]
\centering
\includegraphics[width=0.975\textwidth]{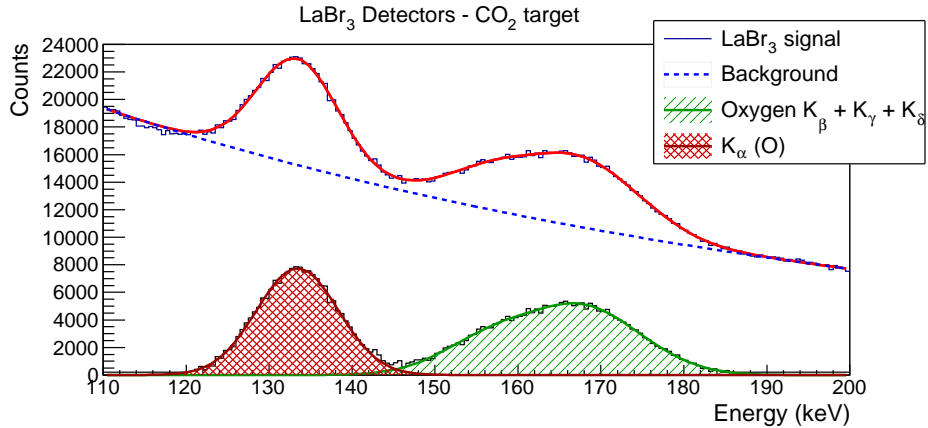}
\caption{The Lyman series of the $\mu$O in the energy spectrum measured by the 1 inch LaBr$_3$ detector with the H$_2$+4$\%$CO$_2$ gas target. The spectrum has been fitted by the composition (red line) of four Gaussian distributions
 (K$_\alpha$, K$_\beta$, K$_\gamma$ and K$_\delta$ lines) and an exponential background.}
\label{fig:co2_labrmi}       
\end{figure}
 In case of LaBr$_3$ detectors two peaks can be clearly distinguished from the background: the first corresponds to the K$_\alpha$ line, the second comes from the sum of the 
 contribution of all higher K lines of $\mu$O. 

\subsection{Time spectra}
The study of the time distribution of the detected signals permits to
 determine not only the lifetime of the muonic atoms but also the
 transfer rate of muons from the $\mu$p to other elements once the
 delayed X-ray muonic transition lines can be selected. Precise time
 spectra can be obtained both using the germanium and the LaBr$_3$ detectors. 

\subsubsection{Germanium detectors}
Thanks to their high energy resolution, HPGe detectors show very
 stable and reproducible signal shape at a specific photon energy. In this condition the time jitter of the acquire data for a specific energy line  is very small.

To study the time resolution of the HPGe detector, a preliminary
 analysis of the time distribution of events in the H2+4\%CO2 gas
 target was performed. 
 Using the high energy resolution of the HPGe detector it
 was possible to select a specific photon energy and reconstruct the
 time distribution of the events contained in the detected energy
 peak. 

\begin{figure}[!tb]
\centering
\includegraphics[width=0.9\textwidth]{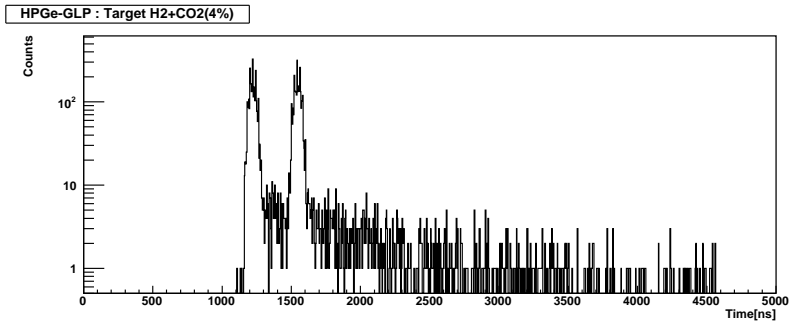}
\caption{Time distribution of events in the oxygen K$\alpha$ spectral
  line at 133 keV as measured by the HPGe-GLP detector.}
\label{fig13}       
\end{figure}

The oxygen K$\alpha$ spectral line at 133 keV shows a FWHM energy
 resolution of 0.89$\pm$0.02 keV measured using a 0.5 microsecond shaping
 time. By selecting events contained inside two times the FWHM energy
 resolution around the oxygen K$\alpha$  peak it was possible to
 reconstruct the time spectra reported in Fig.~\ref{fig13}.
 The two peaks correspond to the
 muon beam double pulse structure, separated by 321.1$\pm$0.3~ns
 demostrating the high precision of differential time measurements of
 the HPGe-GLP detector. 
 Events presently measured in the long tail are statistically compatible with the background of the measurements due to energy released in the detector by means of environmental photons of produced by electron bremsstrhalung. A net increase of the statistics may help very much in the identification of real delayed signal events over the background.
 The time distribution has an offset of about 1100~ns respect to the LaBr$_3$ detectors
 since the delay due to the trigger cable connected to the digitizer
 has not been subtracted in this plot.

These observations lead to conclude that, despite the poor time
 resolution on a single event recorded by the HPGe detector due to the
 long shaping time, the differential time measurements is extremely
 precise and shows time resolution of the order of few nanoseconds. At
 this stage the HPGe detector demonstrated to be suitable to 
 select the specific photon energy produced by the muonic atomic
 transition and permitting, at the same time, to achieve an excellent time distribution of such well identified photons. In this way, and with more statistics, it will be possible to study in details the time behavior of the muon capture and the muon transfer inside the gas target.

\subsubsection{LaBr$_3$ detectors}
LaBr$_3$ detectors were chosen due to their fast light signal. The
 LaBr$_3$ mosaic detector was placed far enough from the target to
 minimize pile-up effects and, for this reason, it was preferred for these studies. When high-energy photons or charged particles signal saturates, the detector can remain ``blind'' for several nanoseconds. Since beam is pulsed and the muons decay exponentially with a life-time of about 2200 ns, the effects of saturation are time dependent and are stronger in coincidence of the beam pulses. Saturation ``blind'' time as function of time from the start of the beam pulses was evaluated on an event-by-event basis and an average saturation correction function was determined for each run. Saturation corrections were properly taken into account for all the timing data used in this section.

\begin{figure}[!tb]
\centering
\includegraphics[width=0.9\textwidth]{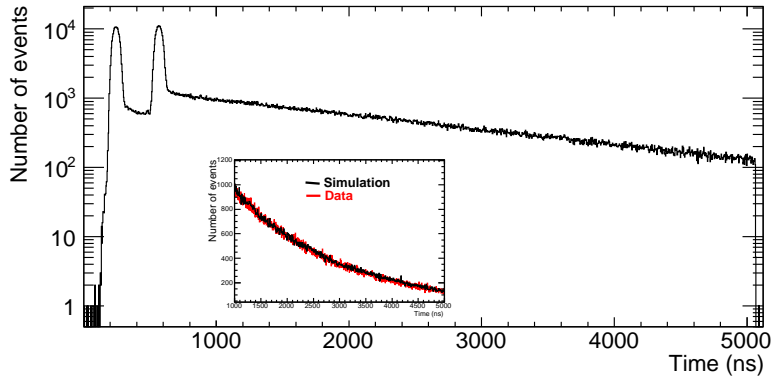}
\caption{Time distribution of the signals (integral of energy spectrum) coming from the LaBr$_3$ mosaic detector for the graphite target. Inset: details of the tail, comparison between simulation (black) and data (red). }
\label{fig:t_c}       
\end{figure}

Figure \ref{fig:t_c} 
 shows the time distribution of the signals detected over the whole energy spectrum by the LaBr$_3$ mosaic detector using the graphite target. The two intense peaks with FWHM $\approx$ 70 ns separated by 320 ns corresponds to the two beam spills. In fact, as soon as the muons are captured by the target materials they emit the typical muonic atoms X-rays (prompt emission). A long tail follows the two peaks, the signals recorded are coming both from electrons (with energy of about 50 MeV) coming from the muon decay and from bremsstrahlung photons (a continuous exponential tail from few keV to the MeV region) emitted by these electrons. A GEANT4 simulation~\cite{agostinelli} of the target confirms this hypothesis, see inset in Fig. \ref{fig:t_c}. In the X-ray energy range of interest (some hundreds of keV) the bremsstrahlung photons are the main source of background. Nevertheless, the time distribution of the tail depicts correctly the muon life time inside the target.

Due to the double Gaussian pulse beam and due to the exponential decay
 of muonic atoms, the shape of the tail can be analytically represented
 by the sum of two Gaussian convoluted with exponential functions. The
 two Gaussians represent the double pulse structure of the muon beam
 while the exponential lifetime parameter is the same for all the exponential
 functions. Using this parametric representation it was possible to
 fit the ta il of the graphite target distribution in order to obtain the lifetime $\tau_C$ of the $\mu$C atom, which resulted to be $\tau_C = 2011\pm16$ ns. This result is in agreement with the expected value of $2026.3\pm1.5$ from reference \cite{suzuki}.

\begin{figure}[!tb]
\centering
\includegraphics[width=0.9\textwidth]{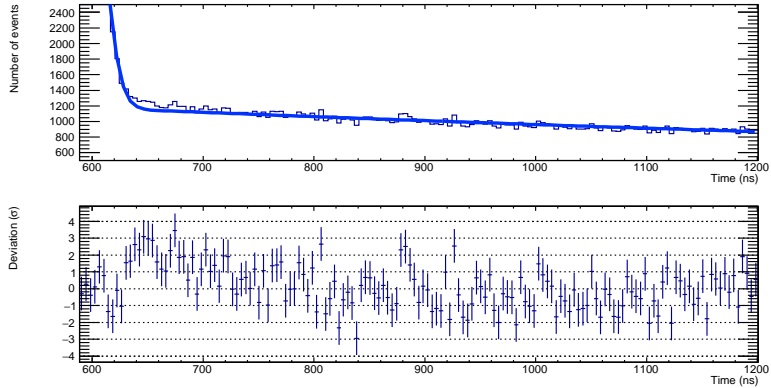}
\caption{Top panel: fit of the time distribution for the graphite target, close up of the region after the second pulse. Bottom panel: deviation of the data from the fit ($\sigma$) in the same time region.}
\label{fig:t_c2}       
\end{figure}

Figure \ref{fig:t_c2} 
 shows a close up of the result of the fit in the time range soon after the second pulse (top panel). The fit was performed starting at 580 ns, assuming a Gaussian tail for the distribution of prompt signals and excluding the range from 610 to 2500 ns in which there is the transition from prompt and delayed events -- when using gas mixture targets. The bottom panel shows that the extrapolation of the fit in the excluded region is in a good agreement with data. 

\begin{figure}[!tb]
\centering
\includegraphics[width=0.9\textwidth]{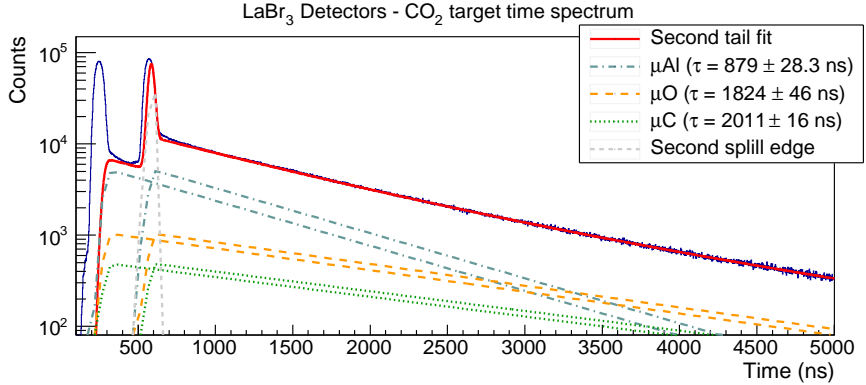}
\caption{Fit of the time distribution for the carbon dioxide target. The fit region goes from 580 to 610 ns and from 2500 to 5000 ns. Aluminium (blue dot-dashed lines), oxygen (orange dashed lines) and carbon (green dotted lines) contribution to the resulting fit curve (red line) are shown.}
\label{fig:t_h2co2tot}       
\end{figure}

The next step was the analysis of the time distribution obtained using the pure H$_2$ target. In this case the overall behaviour is similar to the graphite target but two type of exponential functions are expected: muons interact and stops within both the aluminum container and the gas target. Considering the double pulse, this means that the 
 analytical representation of the distribution is made of four Gaussian convolution of exponentials with two different lifetimes, one representing the $\mu$Al and one representing the $\mu$p atom decays. By fitting the time distribution a resulting value for the $\mu$Al atoms of $\tau_{Al} = 879\pm28$ ns and value for the $\mu$p atoms $\tau_p = 2141\pm98$ ns were obtained, in agreement with expected values from reference \cite{suzuki}. In the case of the  H$_2$+4\% CO$_2$ target, two new exponential functions must be added in comparison to the H$_2$ target analytical description of the time distribution tail. These exponentials represent the $\mu$C and $\mu$O decay time. In this case, only absolute normalizations and the decay rate for oxygen where left free in the fitting procedure, while the life times of already measured atoms were fixed. The fit was performed as before by excluding the time region from 610 to 2500 ns and by assuming that the tail of the second pulse is exclusively made of prompt events. Results are shown in Fig. \ref{fig:t_h2co2tot},
 where the measured life time for the $\mu$O atoms was $\tau_O = 1824\pm46$ ns, compatible with previous measurements \cite{suzuki}. However, the more important result of the fit concerns the normalization of the Gaussian convoluted exponential functions for the various atoms: the contribution to the tail of the $\mu$p decay resulted negligible and compatible with zero, while the $\mu$O and $\mu$C normalization factors resulted to be $1065\pm159$ and $497\pm66$, respectively. The fact that the tail of the time distribution is well represented (the fit converges with $\chi^2/NDF \sim 0.95$) by $\mu$O and $\mu$C only with a ratio compatible with the true atomic concentration of the two elements is a strong indication that the $\mu$p atoms transfer very quickly the muons to the higher charge nuclei.
\begin{figure}[!tb]
\centering
\includegraphics[width=0.9\textwidth]{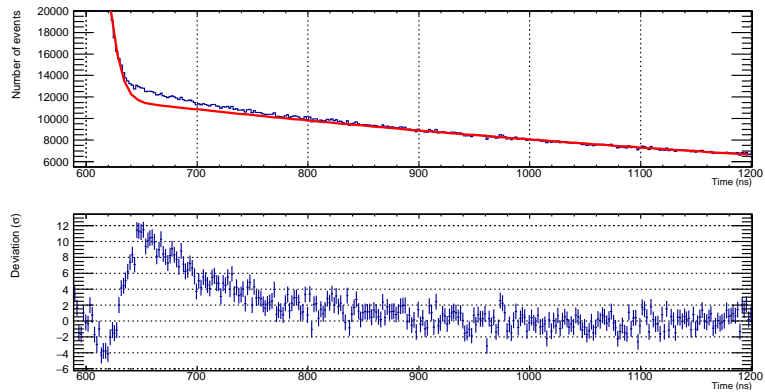}
\caption{Top panel: fit of the time distribution for the carbon dioxide target, close up of the region after the second pulse. Bottom panel: deviation of the data from the fit ($\sigma$) in the same time region. A highly significant deviation of data from the fit due to muon transfer from $\mu$p to oxygen and carbon can be noticed.}
\label{fig:t_h2co2}       
\end{figure}
 Moreover, by comparing the resulting fit to the data in the region
  close to the second pulse (Fig. \ref{fig:t_h2co2}) and also by
  assuming that the pulse rise is completely due to prompt emission, a
  very significant deviation (up to twelve sigmas) from the pure $\mu$O
  and $\mu$C fit emerges. This is the tail of the muonic oxygen X-rays
  emitted when the $\mu$p atoms transfer the muon to the oxygen and
  carbon nuclei. The same effect can be seen also after the first pulse (not shown in figure).

A similar situation was found with the argon target. From the fit of the time distribution tail a value of $\tau_{Ar} = 564\pm14$ ns was measured as $\mu$Ar life time, again in agreement with previous measurements \cite{suzuki}.

From a study of the time evolution of the X-ray events within the oxygen (argon) peak it will be possible to determine the muon transfer rate from hydrogen to oxygen (argon). The results of this analysis will be published in a following paper.

\section{Conclusions}
The RIKEN-RAL double pulsed muon beam has been proved to be an
 excellent source for the FAMU experiment. The first test performed by
 the FAMU collaboration confirmed that it was possible to setup an
 excellent apparatus made of a extremely pure gas target, a beam
 hodoscope and a high resolution X-rays detectors. In such conditions it was not only possible to precisely detect muonic X-rays lines but also to study their time evolution. The recorded data will permit to determine the muon transfer rates from muonic hydrogen to oxygen and to argon.

\section*{Acknowledgements}%\acknowledgments
The research activity presented in this paper has been carried out in the framework of the FAMU experiment funded by Istituto Nazionale di Fisica Nucleare (INFN). The use of the low energy muons beam has been allowed by the RIKEN RAL Muon Facility. We thanks the RAL staff for the help and precious collaboration in the set up of the experiment at RIKEN-RAL port 4.

We gratefully recognize the help of T. Schneider, CERN EP division,
 for his help in the optical cutting of the scintillating fibers of the
 hodoscope detector and the linked problematics and N. Serra from
 Advansid srl for useful discussions on SiPM problematics.


\begin{thebibliography}{dummy}

\bibitem{bakalov92}
D. Bakalov et al., \emph{Experimental method to measure the hyperfine splitting 
of muonic hydrogen ($\mu^-$p)}, {\emph{Phys. Lett. A}} {\bf 172} (1992) 277.
\bibitem{adamczak12}
A. Adamczak et al., \emph{Hyperfine spectroscopy of muonic hydrogen and the PSI 
Lamb shift experiment}, {\emph{Nucl. Instr. Meth. B}} {\bf 281} (2012) 72.
\bibitem{bakalov15}
D. Bakalov et al., \emph{Theoretical and computational study of the energy depen
dence of the muon transfer rate from hydrogen to higher--Z gases}, {\emph{Phys. 
Lett. A}} {\bf 379} (2015) 151.
\bibitem{bib1}
A. Antognini et al., \emph{Proton Structure from the Measurement of 2S-2P Transition Frequencies of Muonic Hydrogen}, {\emph{Science}} {\bf 339} (2013) 417.
\bibitem{bib2}
J.C. Bernauer et al., \emph{High-precision determination of the electric and magnetic form factors of the proton}, {\emph{Phys. Rev. Lett.}} {\bf 105} (2010) 242001.
\bibitem{R512}
A. Adamczak et al., \emph{Measurement of the muon transfer rate from proton to heavier nuclei at epithermal energies}, proposal P484 a Riken-RAL, RAL, UK.
\bibitem{RAL}
T. Matsuzaki et al., \emph{The RIKEN-RAL pulsed Muon Facility}, {\emph{Nucl. Instr. Meth. A}} {\bf 465} (2001) 365.
\bibitem{carbone2015}
R. Carbone et al., \emph{The fiber-SiPM beam monitor of the R484 experiment at the RIKEN-RAL muon facility}, {\emph{J. of Inst.}} {\bf 10/3} (2015) C03007.
\bibitem{branchini2011}
P. Branchini et al., \emph{An FGPA Based General Purpose DAQ Module for the KLOE-2 Experiment} {\emph{IEEE Trans. Nucl. Sci.}} {\bf 58/4} (2011) 1544.
\bibitem{WEBCrio}
www.criotec.com
\bibitem{LABR0}
M. S. Alekhin et al., \emph{Improvement of \u03b3-ray energy resolution of LaBr3:Ce3+ scintillation detectors by Sr2+ and Ca2+ co-doping}, {\emph{Appl. Phys. Lett.}} {\bf 102} (2013) 161915.
\bibitem{LABR1}
R. Pani et al., \emph{Energy resolution measurements of LaBr$_3$:Ce scintillating crystals with an ultra-high quantum efficiency photomultiplier tube}, {\emph{Nucl. Instr. Meth. A}} {\bf 610} (2009) 41.
\bibitem{LABR2}
Saint-Gobain Crystals, \emph{Brilliance 380 Scintillation Material data sheet} and \emph{BrilLanCe$^{TM}$  Scintillators Performance Summary}, www.detectors.saint-gobain.com .
\bibitem{LABR3}
R. B. Firestone and L. P. Ekstrom, \emph{WWW Table of Radioactive Isotopes}, Version 2.1 (2004), http://ie.lbl.gov/toi/index.asp .
\bibitem{LABR4}
B. D. Milbrath, J. I. McIntyre, R. C. Runkle, L. E. Smith, \emph{Contamination in LaCl3:Ce Scintillators}, {\emph{Pacific Northwest National Laboratory}} {\bf PNNL-15453} (2005).
\bibitem{werthmuller96}
A. Werthm\"uller et al., \emph{Transfer of negative muons from hydrogen to oxygen}, {\emph{Hyp. Interact.}} {\bf 103} (1996) 147.
\bibitem{daniele72}
D. F. Measday, \emph{The nuclear physics of muon capture}, {\emph{Phys. Rep.}} {\bf 354} (2001).
\bibitem{schnu92}
H. Schneuwly, \emph{Ephemeral and/or coloured muonic hydrogen atoms}, {\emph{Z. Phys. C - Particles and Fields}} {\bf 56} (1992) S280.
\bibitem{mulhauser93}
F. Mulhauser and H. Schneuwly, \emph{Systematic study of muon transfer to sulphur dioxide}, {\emph{Hyp. Interact.}} {\bf 82} (1993) 507.
\bibitem{werthmuller98}
A. Werthm\"uller et al., \emph{Energy dependence of the charge exchange reaction from muonic hydrogen to oxygen}, {\emph{Hyp. Interact.}} {\bf 116} (1998) 1.
\bibitem{agostinelli}
S. Agostinelli et al., \emph{Geant4 -- a simulation toolkit}, {\emph{Nucl. Instr. Meth. A}} {\bf 506} (2003) 250.
\bibitem{suzuki}
T. Suzuki, D. F. Measday, J. P. Roalsvig, \emph{Total nuclear capture rate for negative muons}, {\emph{Phys. Rev. C}} {\bf 35} (1987).
\end{thebibliography}
\end{document}